# Suppressed Leidenfrost phenomenon during impact of elastic fluid droplets

**Purbarun Dhar**[1, a, *], **Soumya Ranjan Mishra**[2, #], **Ajay Gairola**[2, #], and **Devranjan Samanta**[2, b, *]

[1] Department of Mechanical Engineering, Indian Institute of Technology Kharagpur, Kharagpur–721302, West Bengal, India

[2] Department of Mechanical Engineering, Indian Institute of Technology Ropar, Rupnagar–140001, Punjab, India

*<u>Corresponding authors</u>

[a] E–mail: purbarun@mech.iitkgp.ac.in

[a] Tel: +91–3222–28–2938

[b] E–mail: devranjan.samanta@iitrpr.ac.in

[b] Tel: +91-1881-24-2109

[#] *Equal contribution authors*

**Abstract**

The present article highlights the role of non-Newtonian (elastic) effects on the droplet impact phenomenology at temperatures considerably higher than the boiling point, especially at or above the Leidenfrost regime. The Leidenfrost point (LFP) was found to decrease with increase in the impact Weber number (based on velocity just before the impact) for fixed polymer (Polyacrylamide, PAAM) concentrations. Water droplets fragmented at very low Weber numbers (~22), whereas the polymer droplets resisted fragmentation at much higher Weber numbers (~155). We also varied the polymer concentration and observed that till 1000 ppm, the LFP was higher compared to water. This signifies that the effect can be delayed by the use of elastic fluids. We have showed the possible role of elastic effects (manifested by the formation of long lasting filaments) during retraction in the improvement of the LFP. However for 1500 ppm, LFP was lower than water, but with similar residence time during initial impact. In addition, we studied the role of Weber number and viscoelastic effects on the rebound behaviour at 405º C. We observed that the critical Weber number till which the droplet resisted fragmentation at 405º C increased with the polymer concentration. In addition, for a fixed Weber number, the droplet rebound height and the hovering time period increased up to 500 ppm, and then decreased. Similarly, for fixed polymer concentrations like 1000 and 1500 ppm, the rebound height showed an increasing trend up to certain a certain Weber number and then decreased. This non-monotonic behaviour of rebound heights was attributed to the observed diversion of rebound kinetic energy to rotational energy during the hovering phase.

**Keywords:** *Droplet; Leidenfrost effect; elastic fluids; heat transfer; non-Newtonian fluids; phase change*

# 1. Introduction

At temperatures well above the boiling point of a liquid, droplets are known to levitate above the heated surface. Such an event occurs due to the rapid formation of a vapour cushion beneath the droplet, over which it hovers and levitates without direct contact with the surface. This phenomenon is known as the Leidenfrost effect, and was first observed by J. G. Leidenfrost in 1751 [1, 2]. The thin vapor layer responsible for the levitation of the droplet has poor thermal conductivity and can significantly hamper the heat transfer process between the solid and the liquid, despite being in the phase change regime. The Leidenfrost effect can have important implications in various applications like thermal management of nuclear power stations, spray cooling, fire suffocation and spray quenching. From a utilitarian perspective, it is desired that the Leidenfrost effect be suppressed, in order to ensure safety and reliability of components.

Since a considerable amount of research done on the Leidenfrost effect using Newtonian fluid droplets has been summarized in various articles [3-4], we will only discuss briefly about the more important aspects of droplet Leidenfrost phenomena. Majority of such discoveries have been due to the rapid strides made in experimental techniques in the last two decades, like high-speed photography, interferometry and surface engineering. Impact dynamics of droplets on superheated surfaces have been categorized as "contact boiling", "gentle film boiling" and "spraying film boiling" [5], and the conditions under which these phenomena occur have been experimentally determined. At comparatively lower temperatures, the droplet stays in contact with the hot surface and undergoes boiling ("contact boiling"), followed by "gentle film boiling" where the droplets form an immediate vapor layer beneath it and bounces off the surface. The simultaneous existence of the vapour layer and ejection of tiny droplets during boiling has been termed as "spraying film boiling". The relevant time and length scales of droplet Leidenfrost dynamics have been studied by the same group [6]. Once the temperature exceeded the Leidenfrost point (LFP), the droplets were observed to undergo directed self-propulsion on ratcheted surfaces [7-8], as well as random self-propulsion on substrates without any surface engineering [9].

On ratcheted superhydrophobic (SH) surfaces, the vapour cushion beneath the droplet was found to be more stable [10], and the droplets exhibited self-propulsion at temperatures below the boiling point (~ 80°C). Using interferometry, the thickness of the vapour cushion in the Leidenfrost regime was measured and was found to be of concave shape [11]. Further, the vapour cushion was found to be dependent on drop shape but not on the substrate temperature. In addition to traditional heat transfer utilities, Leidenfrost vapour cushion can be utilized as a drag reduction agent during movements of bluff bodies through a fluid medium [12]. It has also been shown that droplets on textured SH surfaces under low pressure conditions can self-propel and exhibit trampoline-like behavior [13]. With successive collisions, the droplets accelerate and the coefficient of restitution was measured to be greater than one, apparently violating the second law of thermodynamics. The anomaly is caused by the over–pressure below the drop caused by rapid vaporization at low ambient pressure and restricted movement of the resultant vapour cushion due to substrate adhesion and surface texture.

In comparison to Newtonian fluid droplets, the Leidenfrost effect in non-Newtonian fluid droplets has received far less attention [14-19]. The idea of Leidenfrost effect in non-Newtonian fluid droplets rests on fundamental as well as utilitarian implications. From the fundamental point of view, the Leidenfrost effect of such droplets poses several rich insights on the hydrodynamics and associated thermal transport of viscoelastic or elastic fluids. From the applications perspective, such different hydrodynamic effects may be employed to tune the Leidenfrost behaviour of otherwise innate to Newtonian fluid droplets. Upon addition of flexible chain polymers like polyethylene oxide (PEO) or polyacrylamide (PAAM) to water, the droplets exhibited inhibition of splash phenomena and the formation of secondary droplets/atomization during the contact boiling phase [14]. The non-Newtonian fluid droplets also show slightly reduced maximum spreading diameter and rebound to a greater height compared to water droplets for the same impact Weber number *(We)* [15]. Reports also showed that the ratio ($D_{max}/D_o$) of maximum spreading diameter ($D_{max}$) to the initial droplet diameter ($D_o$) scales as ~$We^{0.5}$ and ~$We^{0.25}$ at low and high *We,* respectively [14].

The present article reports and aims to highlight the role of polymer concentration and Weber number on the impact dynamics of non-Newtonian (Boger or elastic fluids) droplets on heated surfaces, with the surface temperature being varied over a wide spectrum to accommodate the inception of Leidenfrost effect and beyond. We first discuss the hydrodynamics and associated Leidenfrost phenomena during the impact of elastic fluid droplets, and portray the differences compared to water droplets under the same conditions. We next focus on the role of fluid elasticity (polymer concentrations) at different *We* on the Leidenfrost point of the fluid. We show that at high temperatures, the droplets exhibit trampoline like behaviour after the inception of the Leidenfrost effect. Rotational rebound of the droplets under certain thermo-hydrodynamic circumstances has been observed and discussed. The effect of *We* and viscoelasticity on the rebound heights at these high temperatures has also been explored.

## 2. Experimental materials and methodologies

A schematic of the experimental setup is illustrated in figure 1. The droplets were dispensed using a precision droplet dispenser (Holmarc Opto-mechatronics, India) with a digitized control unit and volumetric accuracy of ± 0.1 µL. The droplets were released from glass chromatography microliter syringes with stainless steel needles (22 gauge). The dispensing needle height can be adjusted to different heights (h) to vary the impact velocity. Assuming free fall and neglecting air drag, the velocity of the droplet before impact was measured as $V = \sqrt{2gh}$. In the present article, we have defined Weber number as $We = \frac{\rho D V^2}{\sigma}$ where ρ is fluid density, D is drop diameter and V is the velocity a drop attains by its release from height h. A high speed camera (Photron, UK) was used for image acquisition. The images were recorded as 3600-4800 fps at 1024 x 1024 pixel resolution. The experiments were conducted with aqueous solutions of polyacrylamide (PAAM, procured from Sigma Aldrich) of molecular weight ~ 5 million g/mol.

Homogenous polymeric solutions were prepared using a magnetic stirrer for over 2 hours. Solutions of polymer concentrations ranging from 100 to 1500 ppm were used for the different experiments. A rotational rheometer (Anton Paar) with parallel plate geometry of the measurement system was used to measure the rheological parameters like viscosity, and viscoelastic moduli. The relaxation time of the polymeric fluids was deduced employing standard relationships [25].The rheological results indicate that these fluids are typical Boger or elastic fluids [25]. Droplets of diameter ~2.8 mm were released from the syringes to impact on a temperature controlled hot plate arrangement. The droplet diameter is chosen such that it is just less than the capillary length scale for water (to ensure minimal influence of gravitational forces on the droplet dynamics). The hot plate was a polished copper block, attached to a K-type thermocouple, and controlled by a PID controller and temperature regulator. The tip of the thermocouple is placed 1 mm sub-surface to the point of droplet impact. The thermocouple is used to sense the temperature of the hot plate (maximum rating of 650 ºC) and regulate the power controller to provide a constant temperature surface (accuracy of ~ 1.5 ºC). The experiments were conducted on the hot plate for a temperature range of 175º to 405º C. The region of impact was thoroughly cleaned using acetone before each experiment to avoid aberrations due to surface contamination.

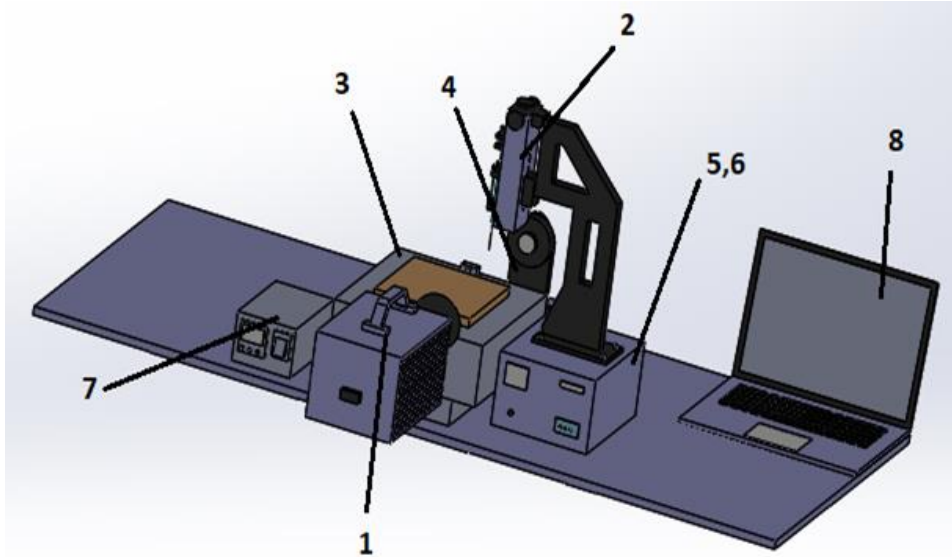

**Figure1:** Schematic of the experimental setup: (1) High speed camera (2) droplet dispenser (3) temperature controlled hot plate (4) LED backlight (5) LED controller (6) Dispenser controller (g) Hot plate controller (h) computer for data acquisition and camera control.

## 3. Results and discussions

**(a) Impact hydrodynamics during the Leidenfrost phenomenon**

The impact hydrodynamics of water droplets on the hot surface at different impact heights was initially studied to probe the effect of different impact *We*. Leidenfrost effect was observed for the first time at 265 ºC, at an impact height of 1.3 cm. We identify the Leidenfrost effect at the surface temperature at which the droplet begins to bounce off the surface. At a relatively lower temperature of 235º C, the water droplets were always in contact with the hot surface (refer fig. S1, supplementary information). Secondary atomization of minuscule droplets was observed. The formation of secondary droplets was due to the rapid contact boiling and vigorous bursting of the vapour bubbles generated at the contact between the droplet and the heated surface [14]. At 265º C, the water droplets were observed to rebound for the first time. Interestingly, ejection of the secondary droplets from the mother droplet was still observed at 265º C. This hints towards a transition regime between the Leidenfrost phenomena and contact boiling, wherein the vapour cushion is potent enough to propel the droplet away from the surface, but at the same time not sufficiently thick to arrest atomization boiling due to rapid heating of the droplet.

At a sufficiently high temperature of 405º C, the droplets were rebounding off the hot surface without ejection of any secondary droplets. This signifies that the vapour cushion is thermally insulating enough to prevent heat transfer (and consequent vapour pocket eruptions) directly into the droplet. The rebound of drops off the wall at high temperatures is due to the formation of a vapour cushion between the hot substrate and the drop. This phenomenon is similar to film boiling and has been termed as the droplet Leidenfrost phenomenon [20, 21]. After the initial investigations with water droplets, experiments were performed with non-Newtonian PAAM solutions. Figure 2 illustrates the comparison of water and aqueous solution of PAAM (500 and 1000 ppm) droplets impacting on hot substrates at 235º C. In comparison to water, the non-Newtonian fluid droplets did not exhibit any secondary atomization, which is in tune with a previous report [14]. The high extensional viscosity of polymer solutions (typical Boger fluids in the present case) is believed to be responsible for suppressing the secondary atomization [14].

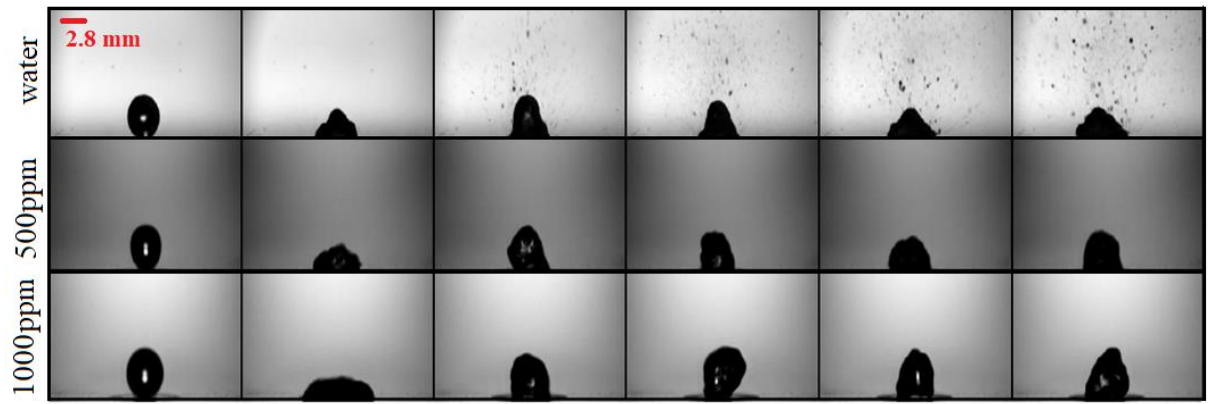

**Figure 2:** 1st row: water; 2nd row: 500 ppm; 3rd row: 1000 ppm of PAAM solution droplets impacting on hot substrate at 235º C, released from 1.3 cm height (We~10) Images are taken at 1.39 milliseconds apart.

Figure 3 illustrates the comparison of dynamics (after maximum spreading) of water and 500 and 1000 ppm PAAM solution droplets, on the hot plate at 405º C, released from 5.1 cm height. At 405º C, the water droplets undergo explosive boiling phase change; undergoing fragmentation and the smaller droplets were observed to rebound. The Leidenfrost rebound is preceded by a period of vigorous spray or explosive boiling. The 500 ppm PAAM solution was observed to generate vertical ligaments/filaments emanating from the bulk portion of the droplet. Small beads are observed to form at the end of the ligaments (fig. 3, 2nd row (ii) and (iii)), which resemble beads-on-a-string structures or BOAS [22-24]. Essentially, explosive or spray boiling is initiated, however, is prevented by the formation of BOAS in the PAAM solution droplet. Eventually, the bulk portion of the droplet rebounds off the surface.

Unlike water droplets, where the thin filaments break up into smaller droplets (and eventual spray formation) due to Rayleigh-Plateau instability, the viscoelastic liquid filaments are stable due to the rapid rise of extensional viscosity with higher strain rates. As the filaments decrease in diameter, enhancement of extensional viscosity retards the capillary driven thinning of the filaments, and further breakup leading to atomization is arrested [22]. With increased polymer concentration, the 1000 ppm solution did not show any atomization or formation of vertical ligaments. It is plausible that the localized flash vaporization within the droplet is arrested by the increased elasticity of the fluid, which manifests as the absence of any vertical ligaments. Just before the rebound event occurs, formation of a very thin filament structure attached to the hot substrate and the upper bulk portion of the droplet was observed (fig. 3, bottom row, vi).

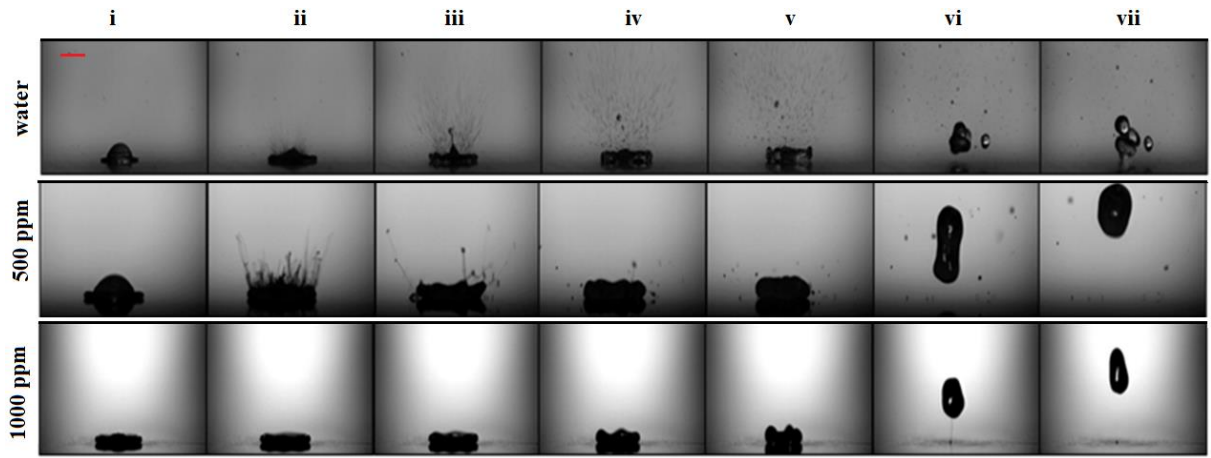

**Figure 3:** Comparison of droplets of water, 500 and 1000 ppm PAAM solution released from 5.1 cm (*We~38*) on hot surface at 405º C (the first 5 photos are 1.39 ms apart, and the time gap between 5$^{th}$-6$^{th}$ and 6$^{th}$- 7$^{th}$ is 11.12 ms).

Similar to water droplets fragmenting at higher impact heights, the polymer droplets also fragment with increasing impact height. Figure 4 shows the impact dynamics of 500 ppm solution droplet released from 20.4 cm (*We~156*) on surface at 405º C. During retraction phase, vertical ligaments are noted to eject out from the bulk of the droplet. However, unlike the case shown in fig. 3, the 500 ppm drop fragmented during the onset of rebound phase (fig. 4, row 1, 4$^{th}$ image). During fragmentation, the smaller daughter droplets were connected to the mother droplet by thin horizontal ligaments or filaments. The longevity of the filaments compared to Newtonian fluid droplets was comparatively higher due to the viscoelastic properties of the 500 ppm PAAM solution. The top views of impact hydrodynamics of droplets (on surface at 405 ºC) of water, 500, and 1000 ppm solution (figure 5) shows the differences in fragmentation behavior at this *We* (~156). The difference between the breakup process of water and 500 ppm solution is visually evident. While the water droplets shatter in several finger-like liquid jets (which later form discrete daughter droplets due to Rayleigh-Plateau instability) coming out from the central part, the 500 ppm solution shows nucleation and proliferation of holes in the film, and subsequent breakage with several filamentous structures connected to each other.

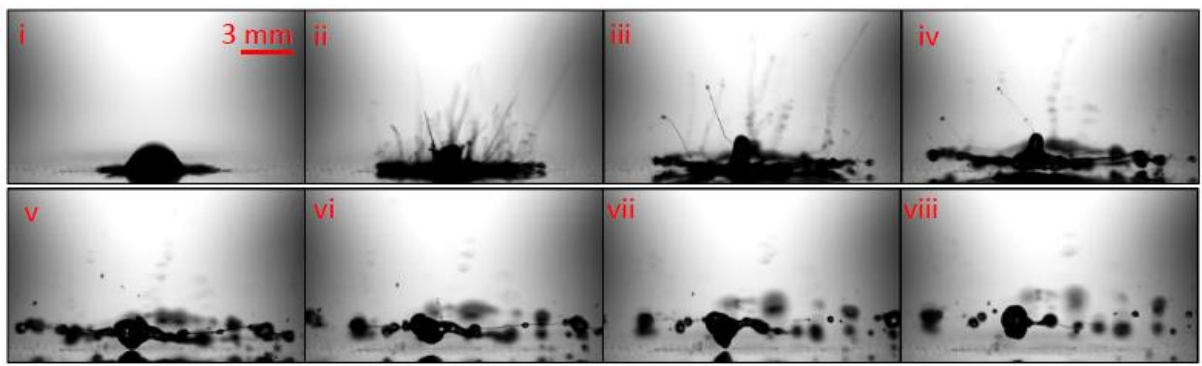

**Figure 4:** 500 ppm solution droplet released from 20.4 cm on to 405º C surface undergoing fragmentation (1.12 ms apart) .The secondary droplets formed at the tip of the ligament

resemble the beads-on–a-string structures (BOAS) reported earlier in previous, albeit dissimilar studies [23].

The breakup process of the polymer drops is retarded by viscoelastic effects and takes nearly twice the time of water droplet. Hence, despite the instability induced by the Leidenfrost phenomenon, the polymer droplets have larger residence times compared to water droplets, even at largely elevated surface temperatures. This signifies that the time available for heat transfer from the surface to the droplet is enhanced by the use of dilute polymer solutions, which may have strong utilitarian implications. With increase of PAAM concentration to 1000 ppm, no fragmentation was observed. The droplets stayed intact and rebounded (fig. 3 bottom row) similar to the lower temperature cases. This is caused by the elongational viscosity of the polymer solutions. When spreading or retracting while hovering on the thin vapor cushion, the droplet undergoes extensional or contractile flow, unlike shear flows during spreading of droplets on surfaces. As the droplet spreads, the filaments start extending from the droplet edge, leading to enhanced extensional shear in the filaments. This leads to stretching of the polymer chains, and enhances the elongational viscosity, which in turn suppresses the fragmentation. In the present case, hydrophobic effect is induced by the vapor cushion, and hence the droplet spreads and recoils on the vapor cushion (this phenomena is often employed in self-levitation of Leidenfrost droplets [7,8], which leads to reduced decay of kinetic energy than on a conventional SH surface. The remnant kinetic energy at the end of retraction phase thus allows for the droplet to be bounced off by the vapor cushion. The noted morphing of hydrodynamics is caused by the elastic nature of the fluids, and the viscosity does not play any major role. This has been established in the previous report by comparing the observations with Newtonian fluids of similar apparent viscosity as the elastic fluids [25].

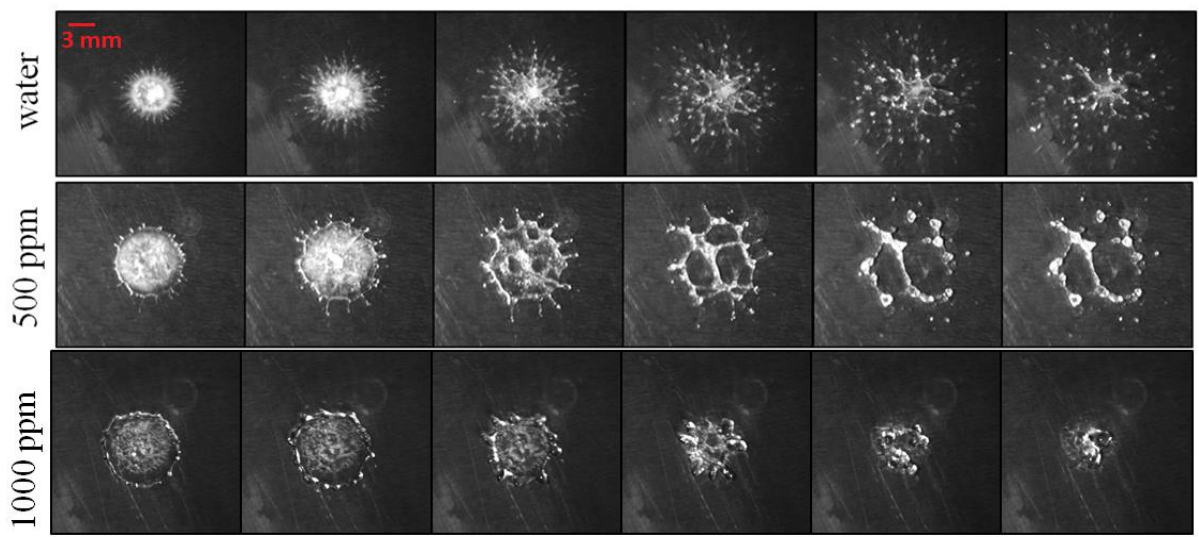

**Figure 5:** Top views of impact Leidenfrost behaviour of (a) water (~0.56 ms apart)  (b) 500 ppm (~1.12 ms apart) and (c) 1000 ppm (~1.68 ms apart) droplets on surface at 405º C, released from a height of 20.4 cm. The fragmentation phenomenon is different in case of non-Newtonian fluids compared to Newtonian fluids (a-b). For 1000 ppm, the droplet it is intact

and undergoes rebound. The side view images of the filamentous structures formed due to onset of elastic instability in polymeric droplets (500 ppm at 405 ºC) have been illustrated in figure S3 (supporting information).

**(b) Effect of fluid elasticity and Weber number ($We$) on the Leidenfrost point (LFP)**

The effect of polymer concentration (which governs the fluid elasticity) and $We$ on the Leidenfrost point (LFP) was probed. The LFP has been defined as the surface temperature at which the droplets show onset of rebound due to levitation by the vapour cushion between the droplet and the substrate. Fig. 6 illustrates the LFP of different polymer concentrations and water. The water droplets stayed intact at lower $We$, and fragmented at higher $We$ at low temperatures. But a reduction in the LFP was noted with increase in the $We$. At higher $We$, the inertia of impact is higher, which leads to enhanced spreading of the droplet. This leads to increased area of contact between the spreading droplet and the hot surface, which increases the effective availability of nucleation sites (at the liquid-solid interface) to form vapour pockets. The enhanced rate of vapour cushion generation thus leads to rebound of the droplet at lower temperature, thereby reducing the LFP. Since the water droplet fragments at higher $We$, the data for water in fig. 6 is restricted to low values of $We$.

In comparison to water, the droplets of polymer solutions stayed intact at higher $We$. At the two lowest $We$, the polymer solutions (100ppm–1000 ppm) exhibited enhancement in LFPs. Also, with increasing impact height and hence higher We, the LFPs showed a decreasing trend, which is in agreement to the mechanism theorized for water droplets in the preceding paragraph. Fig. 6 (b) and (c) illustrate the thin filament formation just before the release of the main polymer solution droplet from the substrate. At 295º C, the 500 ppm solution droplet (fig. 6 b) exhibits filament formation between the two portions of the liquid. This prevented the droplet to rebound fully from the substrate, thereby raising the LFP. Also, the lower columnar part of the droplet (fig. 6b) remained attached to the surface, which is expected to improve heat transfer compared to a traditional Leidenfrost rebound. At lower polymer concentration of 250 ppm (fig. 6c), the droplet rebounded is preceded by the formation of a thin filament and an upper heavy, mushroom-like portion.

With increasing polymer concentration, the overall effectiveness of the droplets in delaying the Leidenfrost effect is noted till ~ 500 ppm. Beyond this, the efficacy deteriorates, and at 1500 ppm, the LFP is lower than water itself (fig. 6a). Fig. S4 (supporting information) illustrates the time of contact or residence time of the droplets between the first impact and rebound from the hot surface, at the corresponding LFP (fig. S4 (a)) and at 405 ºC (fig. S4 (b)). While the LFP of the 1500 ppm case deteriorates below that of water, the residence time of the droplets is same as that of the water droplets. This signifies that while the heat transfer is reduced due to the Leidenfrost effect, the effective thermal transport due to initial contact is not hampered. Overall, from the application viewpoint of elevating the LFP compared to water, the 500 ppm PAAM solution proves to be the most effective.

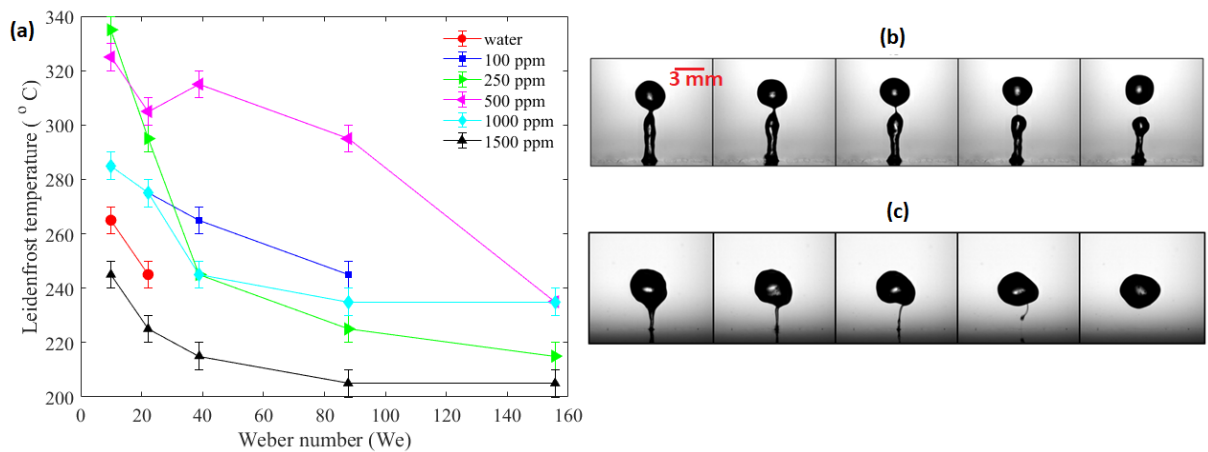

**Figure 6:** (a) Leidenfrost point of droplets with different polymer concentrations at different impact Weber number, (b) 500 ppm PAAM solution released from 2.9 cm on surface at 295º C, shown at 1.39 ms apart (c) 250 ppm PAAM solution on surface at 295º C, shown at 1.12 ms apart. Fig. b and c are of similar scale.

**(c) Influence of fluid elasticity and *We* on the trampoline and rotational dynamics**

At temperatures high enough (~ 405º C), the droplets were observed to rebound repeatedly off the surface and exhibit trampoline-like behavior. Figure 7 (a) illustrates the highest Weber number up to which the various liquid droplets stayed intact without any fragmentation at 405ºC (this is termed as the critical *We*). The critical Weber number was observed to increase with polymer concentration, with a sharp rise after 1000 ppm. Higher the elasticity of the fluid, higher is the propensity of the drop to resist fragmentation. The water (i.e. 0 ppm) droplets fragmented on impact at We >22, whereas the 1500 ppm solution drops survived impacts even at We ~245. Figure 7b illustrates the effect of polymer concentration on the maximum rebound height (observed during the first rebound among the repeated jumps). The maximum rebound height increases up to 500 ppm and then reduces. The retraction velocity of the droplet based on image analysis of (i) 10 successive images from the onset of retraction and (ii) onset of retraction to the moment of drop rebound shows a similar trend as fig. 7 b (fig. S2 in supporting information). With increasing polymer concentration, the elasticity of the fluid increases [26], and the relaxation time, a signature of elasticity is observed to increase with increase in polymer concentration [25]. Hence, for a given Weber number, the droplets with higher polymer concentration possess higher elastic energy, resulting in higher rebound height (fig. 7b) off the vapour cushion.

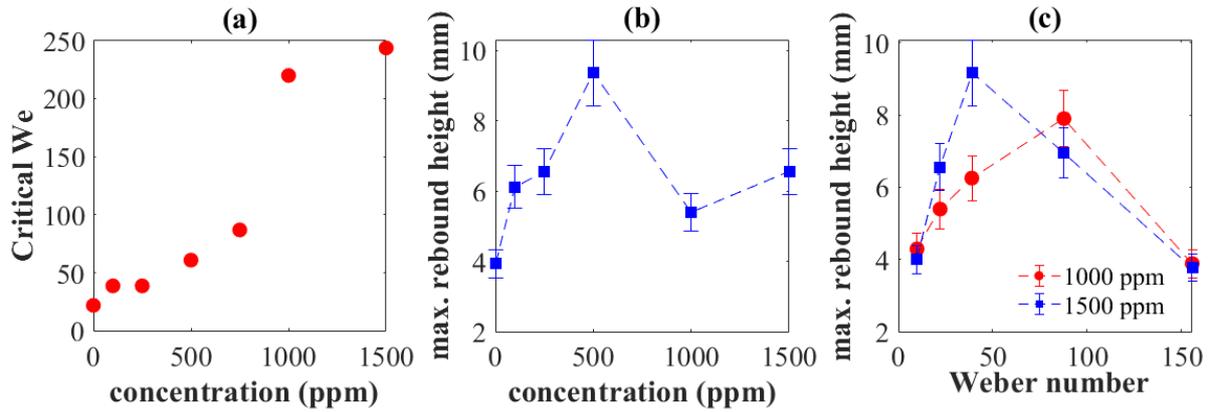

Figure 7: (a) Influence of polymer concentration on the critical Weber number (*We* up to which the droplet stays intact and exhibits trampoline-like behaviour at 405 ºC) (b) the maximum rebound height for the first lift-off for different polymer concentrations, at We~22 and 405 ºC; (We ~22 is selected as it is the highest Weber number at which the water droplet is intact while rebounding at 405 ºC). (c) Maximum rebound height for the first lift-off for 1000 and 1500 ppm solution droplets at different Weber numbers and 405º C.

Figure 7 c illustrates the effect of Weber number on the maximum rebound at 405º C, for the two highest polymer concentrations used in our studies i.e. 1000 and 1500 ppm. Both the curves show a peak value of the rebound height. For a fixed polymer concentration and increasing *We,* the higher maximum spreading resulted in higher stretching of the polymer chains [28]. Therefore, while retraction, the polymer chains recoiled, and the resultant stored elastic energy increased with increasing We. Therefore, in fig. 7c we note an initial increasing trend of rebound height with increasing *We*. However, it was also observed that for both 1000 and 1500 ppm solutions, the rebound height decreases considerably with further increase in *We*. In a previous study [17], it has been reported that the droplets occasionally exhibit somersault like behaviour during the hovering phase post-rebound. The study argued that the diversion of translational kinetic energy to rotational kinetic energy during the somersault like behaviour causes a reduction in the rebound height. Further image analysis of drop motion during hovering of the 1500 ppm solution droplet at 405º C and *We* ~155 showed the somersault like behaviour. Since a part of the translational kinetic energy is converted to rotational kinetic energy responsible for somersault like behaviour, the reduction in the trend of the curves for 1000 and 1500 ppm was observed (fig. 7c). Similar rotational behaviour was observed for We ~88, but no such rotational motion was observed for We~39. Hence, the behaviour is characteristic to impact Leidenfrost effect at higher temperatures and high *We* values. At lower temperatures (the Leidenfrost point of a particular fluid), reduction in hovering time is noted only for the 1500 ppm droplets (refer fig. S5, supporting information).

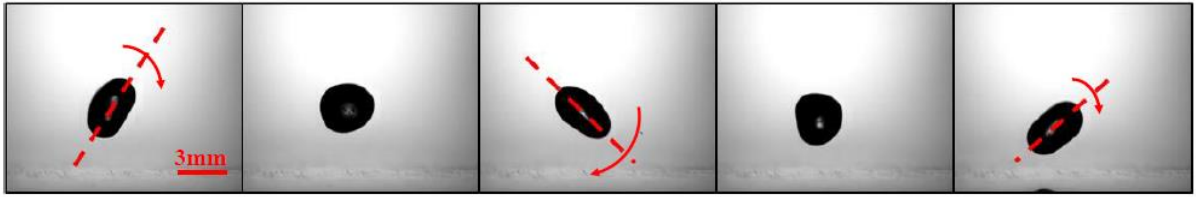

Figure 8: Rotation during the droplet rebound of 1500 ppm solution at 405 °C and We=155. The time interval between successive images is 4 ms. The red dotted line approximates a central axis through the droplet and is placed to emphasize the effect of rotation.

## 4. Conclusions

A recent study on the impact of non-Newtonian fluid droplets [25] showed that once the impact velocity exceeds a critical value, rebound suppression of the droplet is achieved even on superhydrophobic (SH) surfaces. For 1000 and 1500 ppm PAAM solution droplets, the critical velocities were ~2.25 and ~1.75 m/s, respectively (refer table 1 of reference [25]). However on heated hydrophilic surfaces at 405° C, 1000 and 1500 ppm drops were observed to rebound without any fragmentation of satellite droplets at higher velocities of ~2.37 and ~2.50 m/s, respectively (fig. 7a). In previous studies on non-Newtonian fluid droplet impact on SH surfaces, various mechanisms such as the generation of normal stresses, and stretching of flexible chain molecules impeding the receding contact line dynamics were proposed as the reason behind droplet rebound suppression [27-28]. During retraction on SH surfaces, the difference in velocity of the stationary bottom layer attached to the surface and of the top surface generates the shear stress, and induces the elastic effects responsible for rebound suppression.

In the present context, the vapour layer formed at very heated conditions prevents the wetting of the droplet with the substrate and hence the elimination of shear stresses responsible for inhibition of droplet rebound. We believe our result will trigger further interest in the non-Newtonian fluid (specifically Boger fluids) droplet dynamics on heated substrates, especially at and beyond the Leidenfrost regime. We showed that addition of small amounts of long chain polymers could lead to delaying of the droplet Leidenfrost effect. The 500 ppm was found to be optimum polymer concentration for increasing the LFP at different *We*. The elasto-hydrodynamic effects such as filament or ligament formation is noted to arrest associated effects such as vigorous spray boiling, which has a direct role towards delaying the Leidenfrost effect. Although we noted that the 1500 ppm droplets reduced the LFP below water, the residence time of such droplets is similar to water droplets, which is important from heat transfer perspective. For further studies, particle image velocimetry (PIV) measurements can be performed to show the internal circulation within the rebound drops where summersault-like motion is exhibited. Impact behaviour and heat transfer of non-Newtonian fluid droplets on micro-textured or SH surfaces at high temperatures can also be performed along the lines of previous studies [7-11]. Our present findings may have strong implications towards tuning and control of the Leidenfrost effect during droplet and spray based thermal management and allied utilities.

# Suppressed Leidenfrost phenomenon during impact of elastic fluid droplets

**Purbarun Dhar**[1, a, *], **Soumya Ranjan Mishra**[2], **Ajay Gairola**[2], and **Devranjan Samanta**[2, b, *]

**Supporting information**

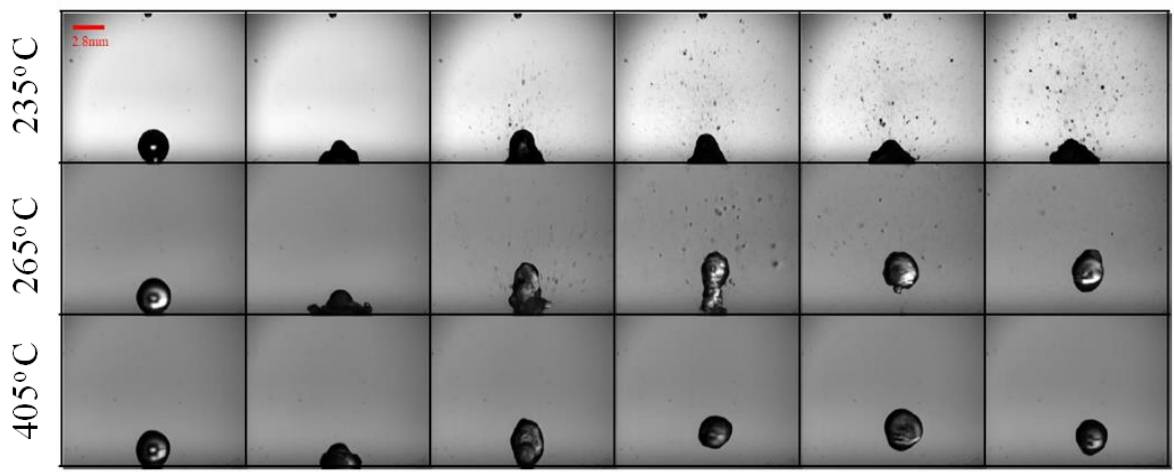

**Figure S1:** Images of water droplet released from 1.3 cm on to different temperature surfaces (top row: drop mostly touches the substrate and ejects spray like secondary droplets (atomization), middle row: after initial atomization, the drop is lifted off from the substrate, and bottom row: at 405°C, the droplet successively rebounds off from the substrate and remains hovering for longer time periods. The behaviour is known as trampoline effect. The images are spaced 8.33 ms apart.

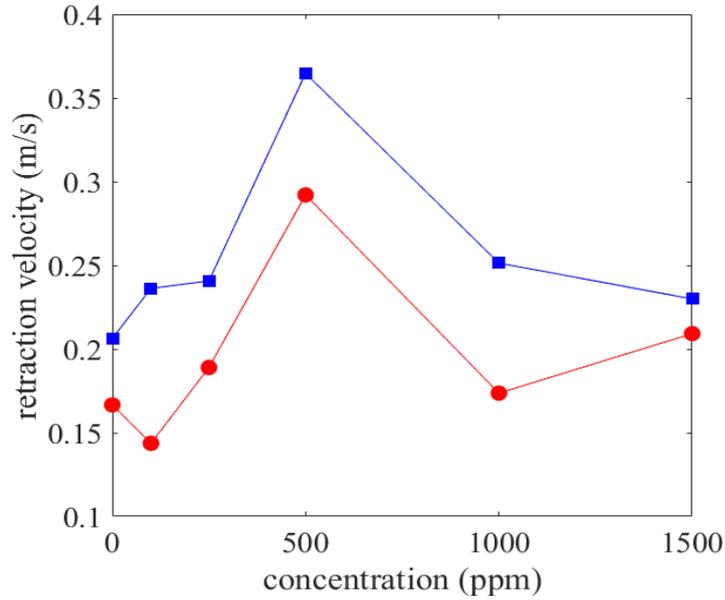

**Figure S2:** Retraction velocity of the droplets of different polymer concentrations at 405 °C and *We* ~22. Retraction velocity is based on image analysis of 10 successive images from (i) from the onset of retraction (red circles) and (ii) from onset of retraction to the moment of rebound (blue squares).

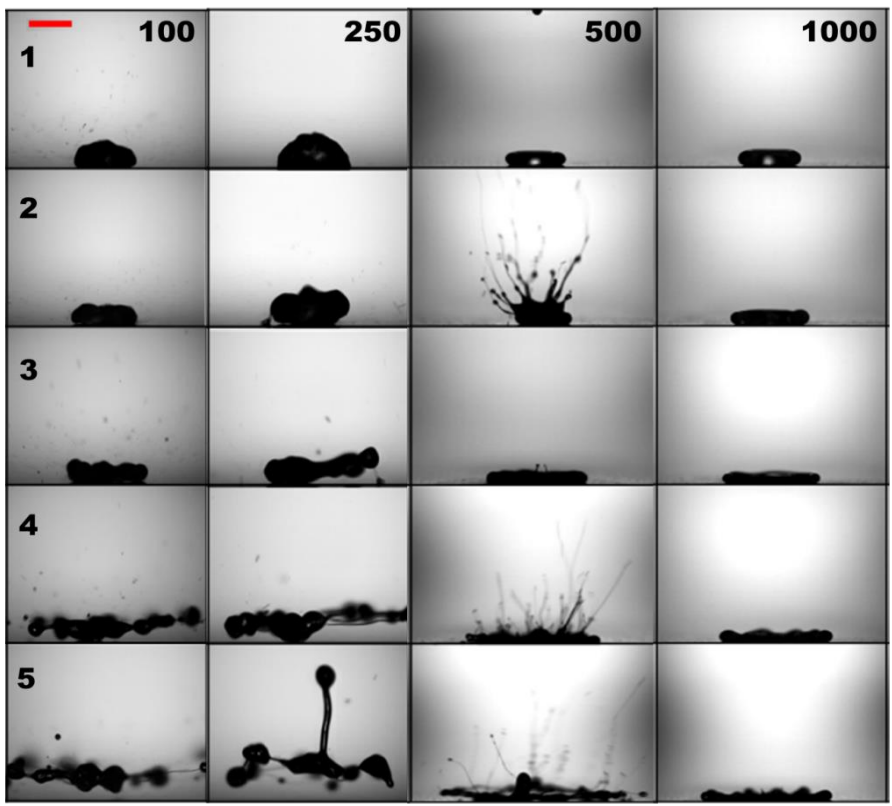

**Figure S3:** Onset of elastic instability and filament dynamics of droplets with different PAAM concentration (the value on the top right of images in the first row represent concentration in ppm for that column) at different heights and at 405 °C. The rows represent impact heights of (1) 1.3, (2) 2.9, (3) 5.1, (4) 11.5, and (5) 20.4 cm. The nature of the filament dynamics at 500

ppm is the most pronounced, with the formation of atypical fungi (250 ppm, last row) or medusoid (500 ppm, 2$^{nd}$ row) structures. Scale (red bar in st row 1$^{st}$ column) represents 2.8 mm. Images are taken at 5.4 milliseconds apart.

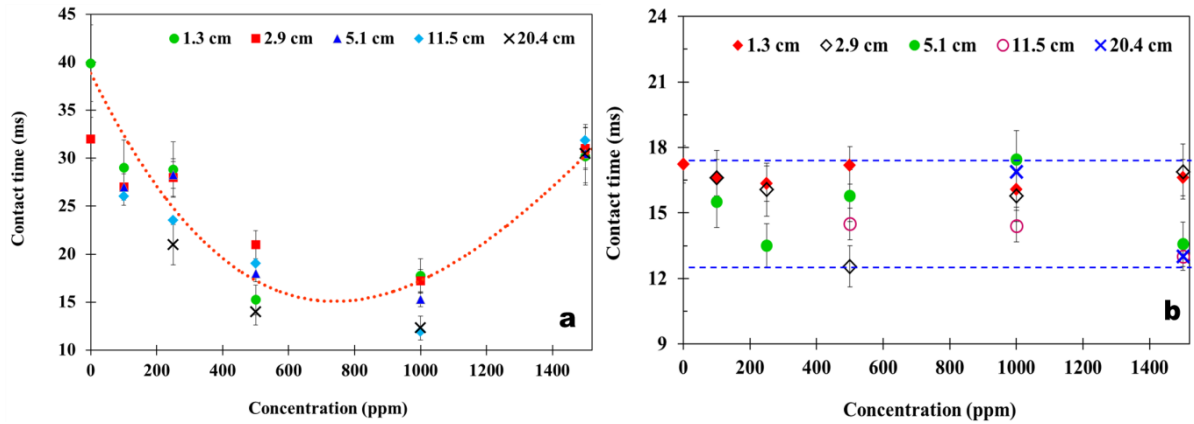

**Figure S4:** Contact or residence time of the droplets between the first impact and rebound, for different polymer concentrations and impact heights, at (a) the corresponding Leidenfrost point and (b) 405 °C. The dotted line in (a) is for guide to the eyes. The dotted lines in (b) represent the band within which all the points lie.

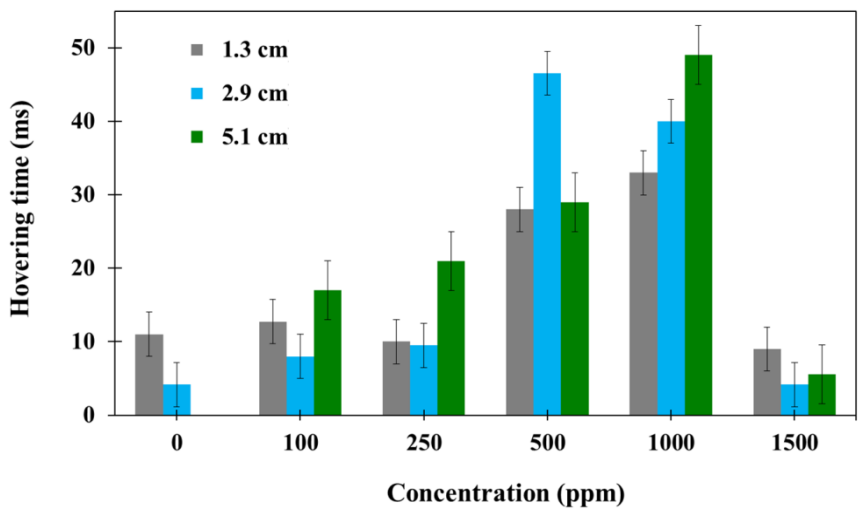

**Figure S5:** Hovering time of the droplet in air after the first rebound at the respective Leidenfrost Point for different polymer concentrations and impact heights.